\documentclass[10pt]{article}
\usepackage{amssymb}
\usepackage{amsmath}
\usepackage{graphicx}
\usepackage{cite}
\usepackage{float}
\usepackage{caption}
\usepackage{authblk}
\usepackage{subfigure}
\usepackage[margin=1.0in]{geometry}
\usepackage{color}

\begin{document}

\title{Universal Behavior of Strain in Quantum Dots}
\author[1*]{\underline{Hesameddin Ilatikhameneh}}
\author[1*]{\underline{Tarek A. Ameen}}
\author[1]{Gerhard Klimeck}
\author[1]{Rajib Rahman}
\affil[1]{\normalsize{Network for Computational Nanotechnology, Department of Electrical and Computer Engineering, Purdue University, West Lafayette, IN 47907, USA}}
\affil[*]{\normalsize{These authors contributed equally to this work.}}
\renewcommand\Authands{ and }
\maketitle
\providecommand{\keywords}[1]{\textbf{\textit{Keywords---}} #1}

\begin{abstract}

Self-assembled quantum dots (QDs) are highly strained heterostructures. the lattice strain significantly modifies the electronic and optical properties of these devices. A \emph{universal behavior} is observed in atomistic strain simulations (in terms of both strain magnitude and profile) of QDs with different shapes and materials. In this paper, this universal behavior is investigated by atomistic as well as analytic continuum models. Atomistic strain simulations are very accurate but computationally expensive. On the other hand, analytic continuum solutions are based onassumptions that significantly reduce the accuracy of the strain calculations, but are very fast. Both techniques indicate that the strain depends on the \emph{aspect ratio} (AR) of the QDs, and not on the individual dimensions. Thus simple closed form equations are introduced which directly provide the atomistic strain values inside the QD as a function of the AR and the material parameters. Moreover, the conduction and valence band edges $E_{C/V}$ and their effective masses $m^*_{C/V}$ of the QDs are dictated by the strain and AR consequently. The universal dependence of atomistic strain on the AR is useful in many ways; Not only does it reduce the computational cost of atomistic simulations significantly, but it also provides information about the optical transitions of QDs given the knowledge of $E_{C/V}$ and $m^*_{C/V}$ from AR. Finally, these expressions are used to calculate optical transition wavelengths in InAs/GaAs QDs and the results agree well with experimental measurements and atomistic simulations.
\end{abstract}


\section{Introduction}

Self-assembled quantum dots (QDs) have improved the performance of many optoelectronic devices, such as quantum dot infrared photodetectors (QDIPs)\cite{liu79,tqd}, intermediate band solar cells (IBSCs)\cite{nextgen,luque2012understanding}, quantum dot optical amplifiers \cite{akiyama2007quantum}, and quantum dot lasers \cite{bimberg1997ingaas,ledentsov2011quantum}; QDIPs  have lower dark current than conventional photodetectors \cite{tqd,ameen2014modeling} and are sensitive to normally incident light unlike their counterparts that are made from quantum wells \cite{tqd,Tarek,Tarek2}. For IBSCs, one of the most successful methods in pushing solar cell efficiency beyond the Shockley-Queisser limit is to add one or more intermediate bands inside the gap, which can be realized by using quantum dots and quantum dot-in-a-well devices\cite{luque2012understanding}. Quantum dot optical amplifiers offer ultra wide band polarization insensitive high-power amplification \cite{akiyama2007quantum}. Quantum dot lasers provide the most reliable and temperature-insensitive operation around the spectral range of 1.15 to 1.25 $\mu$m \cite{ledentsov2011quantum}.

Self-assembled quantum dots are highly strained heterostructures. The atomistic strain in such structures is usually on the order of 10$\%$. The electronic band structure is affected significantly by the lattice strain\cite{ameen2014optimization}. For proper band structure calculations, the strain needs to be calculated and included accurately \cite{boykin2002diagonal}. Such high strain values are beyond the domain of validity of the continuum elasticity theory \cite{pryor1998comparison} and a rigorous atomistic treatment of strain is needed (such as Keating \cite{keating1966} or anharmonic models\cite{laz_2003,ameen2014optimization}). 

Some previous works have calculated strain in QDs using continuum elasticity theory\cite{pearson2000analytical,downes1997simple,andreev1999strain}. These continuum elasticity calculations become inaccurate as the strain exceeds the elastic limit and goes beyond the domain of validity of continuum elasticity, which is the case in these self-assembled QDs.\cite{pryor1998comparison,ameen2014optimization}. In addition, the mathematical modeling problem is complicated and many assumptions are needed to obtain an analytic solution. These assumptions further reduce the accuracy of the analytic solution. Here, these assumptions are discussed and avoided as much as possible for a better accuracy. Nevertheless for a precise treatment of strain, atomistic simulations need to be used. The main issue with atomistic simulations is that they are computationally very expensive. In this work, both approaches are investigated and compared against each other. The analytic solution provides qualitatively similar results to the atomistic simulations. However, it quantitatively underestimates the strain magnitudes. On the other hand, the atomistic simulations of QDs show a universal behavior in the strain results, independent of QD dimensions and materials. Accordingly, compact expressions which capture this universal behavior are put forward as replacements of the analytic solution. These closed form expressions provide the atomistic strain values without the numerical burden of atomistic simulations and the inaccuracy of analytic models.
 
In section II, the analytic continuum solution is discussed. Later on, the atomistic results are shown in section III.  Finally, the universal behavior is introduced and its applications are discussed in section IV.

\section{Analytical Continuum Solution}

In this section, analytic expressions for the strain in cuboid QDs are introduced. These expressions are obtained by solving the continuum elasticity equations \cite{pearson2000analytical} which are based on the stress field Green's function \cite{downes1997simple,faux1996simple,downes1995calculation} used to predict stain in quantum wells, wires, and dots. Unlike previous works \cite{pearson2000analytical, downes1997simple,faux1996simple,downes1995calculation}, hydrostatic strain is not assumed to be a constant here. The assumptions and details of the derivations are provided in the appendix. The strain in the middle of a cuboid quantum dot with base length $b$ and height $h$ is found to be:
\begin{equation}
 \varepsilon_{H}  = \frac{1 - 2 \nu}{E} \left( - 12 \pi \Lambda + 16 \Lambda tan^{-1}\left( \frac{1}{2\sqrt{\left(\frac{b}{h}\right)^2+\frac{1}{4}}} \right) + 8 \Lambda cot^{-1}\left( \frac{2\sqrt{1+\frac{5}{4}\left(\frac{b}{h}\right)^2}}{\left(\frac{b}{h}\right)^2} \right)\right) = f(\frac{b}{h}),
\label{eqn_Aal_1}
\end{equation}

\begin{equation}
 \varepsilon_{B}  = \frac{1 + \nu}{E} 16 \Lambda \left( tan^{-1}\left( \frac{1}{2\sqrt{\left(\frac{b}{h}\right)^2+\frac{1}{4}}} \right) - cot^{-1}\left( \frac{2\sqrt{1+\frac{5}{4}\left(\frac{b}{h}\right)^2}}{\left(\frac{b}{h}\right)^2} \right) \right) = f(\frac{b}{h}),
\label{eqn_Anl_2}
\end{equation}
where $\varepsilon_{H}$ is the hydrostatic strain, $\varepsilon_{H}=\varepsilon_{xx}+\varepsilon_{yy}+\varepsilon_{zz}$, and $\varepsilon_{B}$ is the biaxial strain, $\varepsilon_{B}=\varepsilon_{xx}+\varepsilon_{yy}-2\varepsilon_{zz}$.  $\Lambda = \frac{\varepsilon_0 E}{4\pi (1-\nu)}$, where $\varepsilon_0$ is the lattice misfit, $\varepsilon_0 = \frac{a_{dot}-a_{substrate}}{a_{dot}}$, and $\nu$ and $E$ are the Poisson ratio and Young's modulus of the quantum dot material. The choice of describing the strain in terms of $\varepsilon_{H}$ and $\varepsilon_{B}$ components instead of Cartesian components $\varepsilon_{xx}$, $\varepsilon_{yy}$ and $\varepsilon_{zz}$ is due to the fact that in zinc-blend materials the deformation in band edges depends on the $\varepsilon_{H}$ and $\varepsilon_{B}$ directly.

\begin{figure}[H]
\centering
\includegraphics[width=60mm]{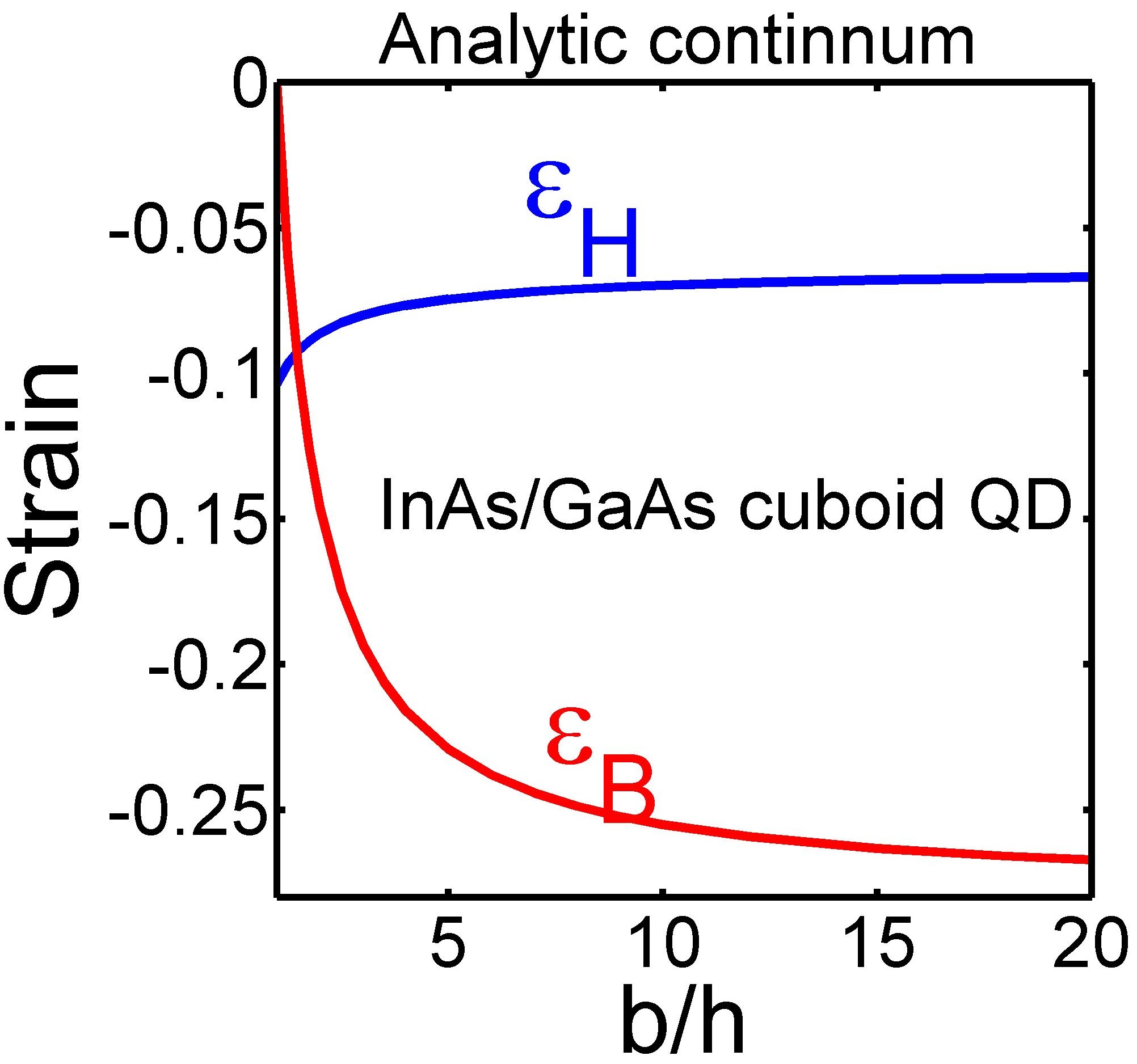}
\caption{The analytic continuum solution of hydrostatic and biaxial strain as functions of b/h. Strain depends on b/h, but not on the individual dimensions.}
\label{HB_analytical_fig}
\end{figure}

The analytic continuum strain solution indicates that the strain depends on the base to height aspect ratio (b/h). Figure \ref{HB_analytical_fig} shows $\varepsilon_{H}$ and $\varepsilon_{B}$ as a function of b/h for a cuboid QD obtained from equations (\ref{eqn_Aal_1}) and (\ref{eqn_Anl_2}). The magnitude of $\varepsilon_{H}$ decreases slightly with increasing b/h, while the magnitude of $\varepsilon_{B}$ increases with increasing b/h.

Notice that the solution in \cite{pearson2000analytical} is based on a stress field Green's function that has been introduced for quantum wires\cite{downes1995calculation}, in which \emph{at least one strain component is assumed to be known}. This assumption is suitable for quantum wires and quantum wells but not for quantum dots.  The main reason behind the deviation from the atomistic simulations of the QDs is this assumption. It is shown in the next section that the analytic solution converges to the atomistic results at large b/h values since the QD becomes more like a quantum well in which this assumption is more justified.


\section{Atomistic Simulation.}

\begin{figure}[H]
\centering
\includegraphics[width=5pc]{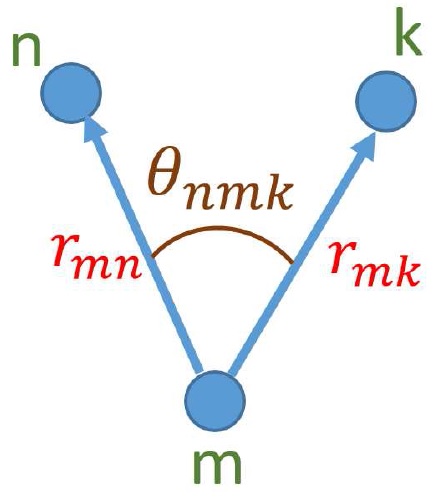}
\caption{Bond lengths and bond angle for three neighbor atoms m, n, and k.}
\label{interactions_fig}
\end{figure}

For an accurate calculation of the strain, a rigorous atomistic strain model is needed to calculate the relaxed atom positions. The Keating model \cite{keating1966} provides the total elastic strain energy of the system as a function of atom positions. 
\begin{equation}
E=\frac{3}{8}\sum\limits_{m,n}{\left[ \frac{{{\alpha }_{mn}}}{d_{mn}^{2}}{{\left( r_{mn}^{2}-d_{mn}^{2} \right)}^{2}}+\sum\limits_{k>n}{\frac{{{{\beta }_{mnk}}}}{{{d}_{mn}}{{d}_{mk}}}\left( {{r}_{mn}}\cdot {{r}_{mk}}-{{d}_{mn}}\cdot {{d}_{mk}} \right)} \right]} , 
\label{eqn_1}
\end{equation}
where the coefficient $\alpha$ corresponds to the force constant for the bond length distortion, and $\beta$ corresponds to the bond angle distortion as shown in Figure \ref{interactions_fig}. $\alpha$ and $\beta$ are material constants. $r_{mn}$ is a vector from atom m to atom n for the strained crystal, while $d_{mn}$  is the same vector for the bulk unstrained crystal. These atomistic interactions are assumed to be between the nearest neighbors only.  

Keating, in his seminal work \cite{keating1966}, showed that there is a straightforward connection between the atomistic force constants $\alpha$ and $\beta$ and the macroscopic elastic constants (e.g. $c_{11}$ and $c_{12}$) of the material. The Keating model is known to be a suitable model for calculation of the atomistic strain in self-assembled QDs \cite{klimeck2007atomistic,klimeck2002development,stier1999electronic,ameen2014optimization} and phonon dispersion in nanowires and in bulk \cite{boykin}. 

\begin{figure}[H]
\centering
\includegraphics[width=12pc]{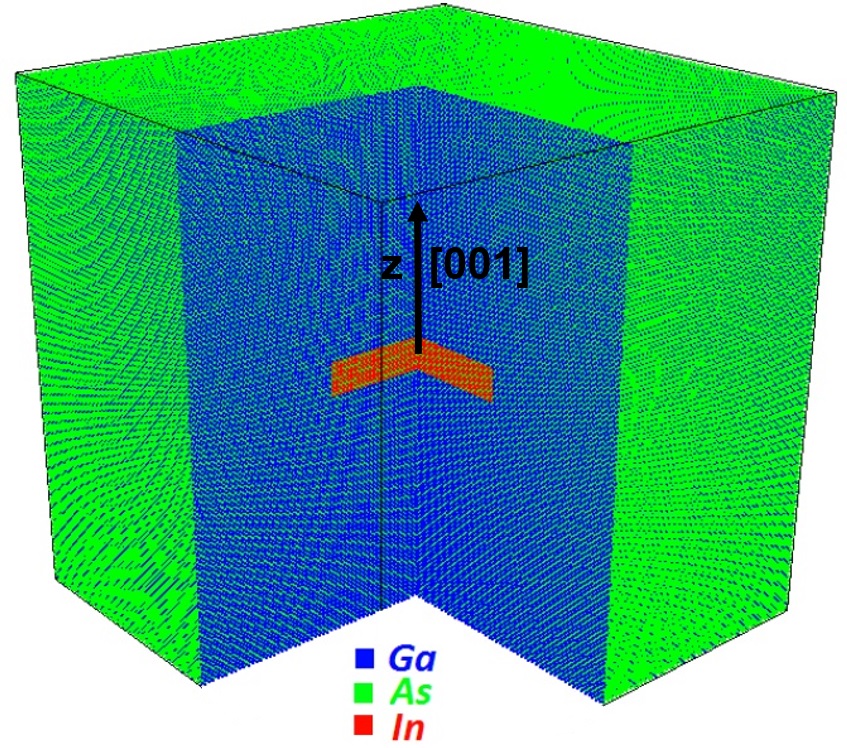}
\caption{The atomistic grid of a cuboid InAs/GaAs QD with the base length of 24 nm and height of 4 nm.}
\label{Atomistic_grid_fig}
\end{figure}

Cuboid and dome shaped QDs with different dimensions have been simulated using the Keating model to study the behavior of strain with different QD dimensions, shapes, and materials. The dimensions of the whole simulated domain are 60 nm x 60 nm x 60 nm and the quantum dot is located in the middle of the structure. The strain simulation contains about 10 million atoms with the atomistic grid shown in Figure \ref{Atomistic_grid_fig}.  Such large systems are computationally expensive for strain simulations and require a highly scalable parallel code. All the atomistic strain simulations in this work have been done by NEMO5 \cite{nemo5, nemo5_2}.

\begin{figure}[H]
\centering
\includegraphics[width=20pc]{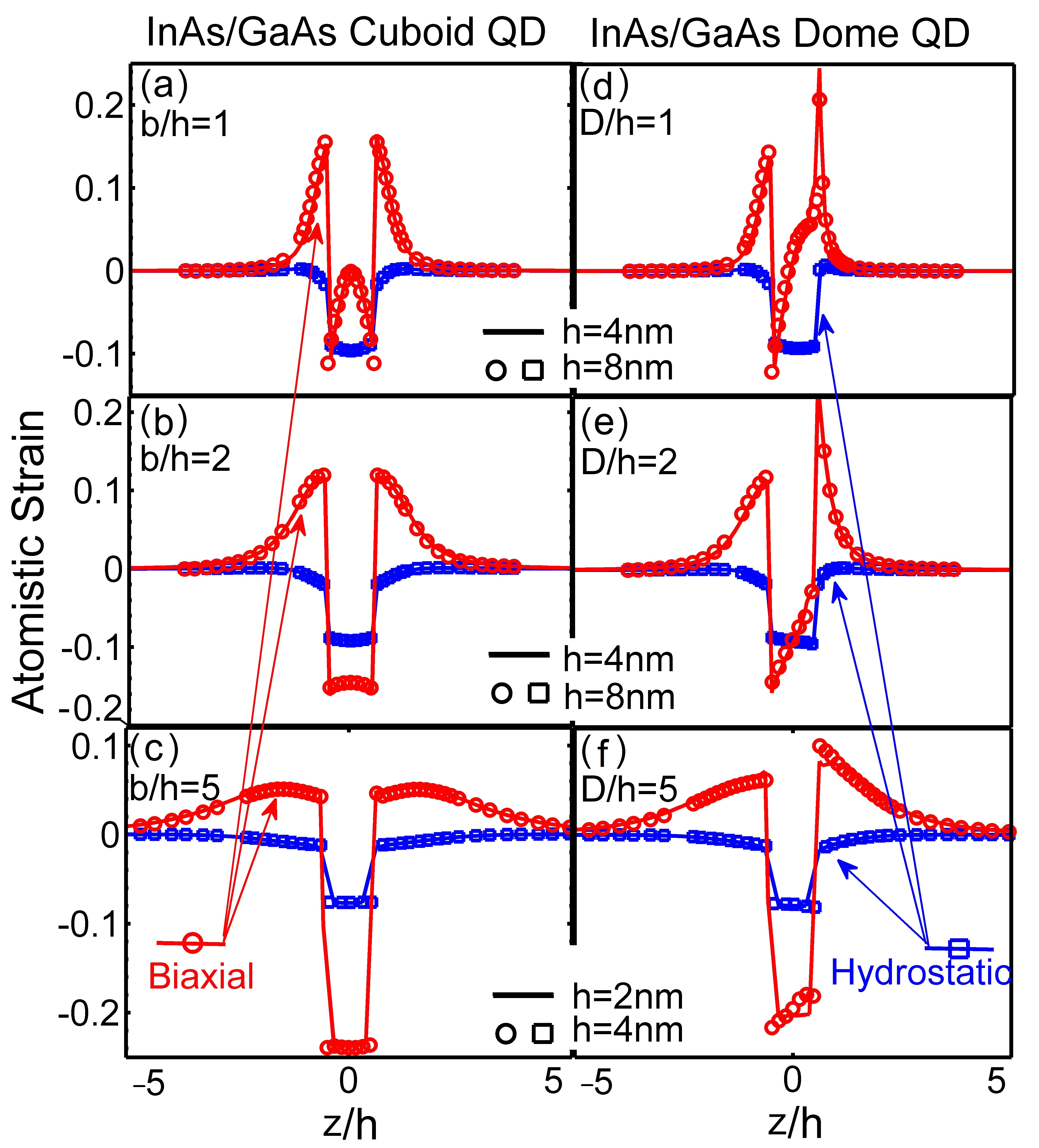}
\caption{Atomistic $\varepsilon_{H}$ (blue) and $\varepsilon_{B}$ (red) strain profiles along a line in the middle of the QD  with the heights of 2nm (lines) and 4nm (symbols) in [001] direction. The solid lines and symbols overlay each others. QDs with different dimensions but same aspect ratios exhibit same strain profiles over the normalized coordinate (z/h) for different aspect ratios (a)-(f). The left three figures are strain distributions of cuboid QDs with the base to height ratios (b/h) of 1 (b), 2 (b), and 5 (c), while the right three figures are for the dome shaped QDs with the base diameter to height ratios (D/h) of 1 (d), 2 (e), and 5 (f). The main point is that, QDs with certain shape and aspect ratio have a unique strain distribution.}

\label{distribution_fig}
\end{figure}

Figure \ref{distribution_fig} shows atomistic $\varepsilon_{H}$ and $\varepsilon_{B}$ along a line in the middle of the device in [001] direction. The positions of the atoms on the line are divided by the dot height (h) for comparison purposes. QDs with the different dimensions but the same aspect ratio have similar strain distribution over the normalized coordinates. This behavior is observed in both dome and cuboid QDs.

\begin{figure}[H]
\centering
        \subfigure[]{\label{fig:a}\includegraphics[width=53mm]{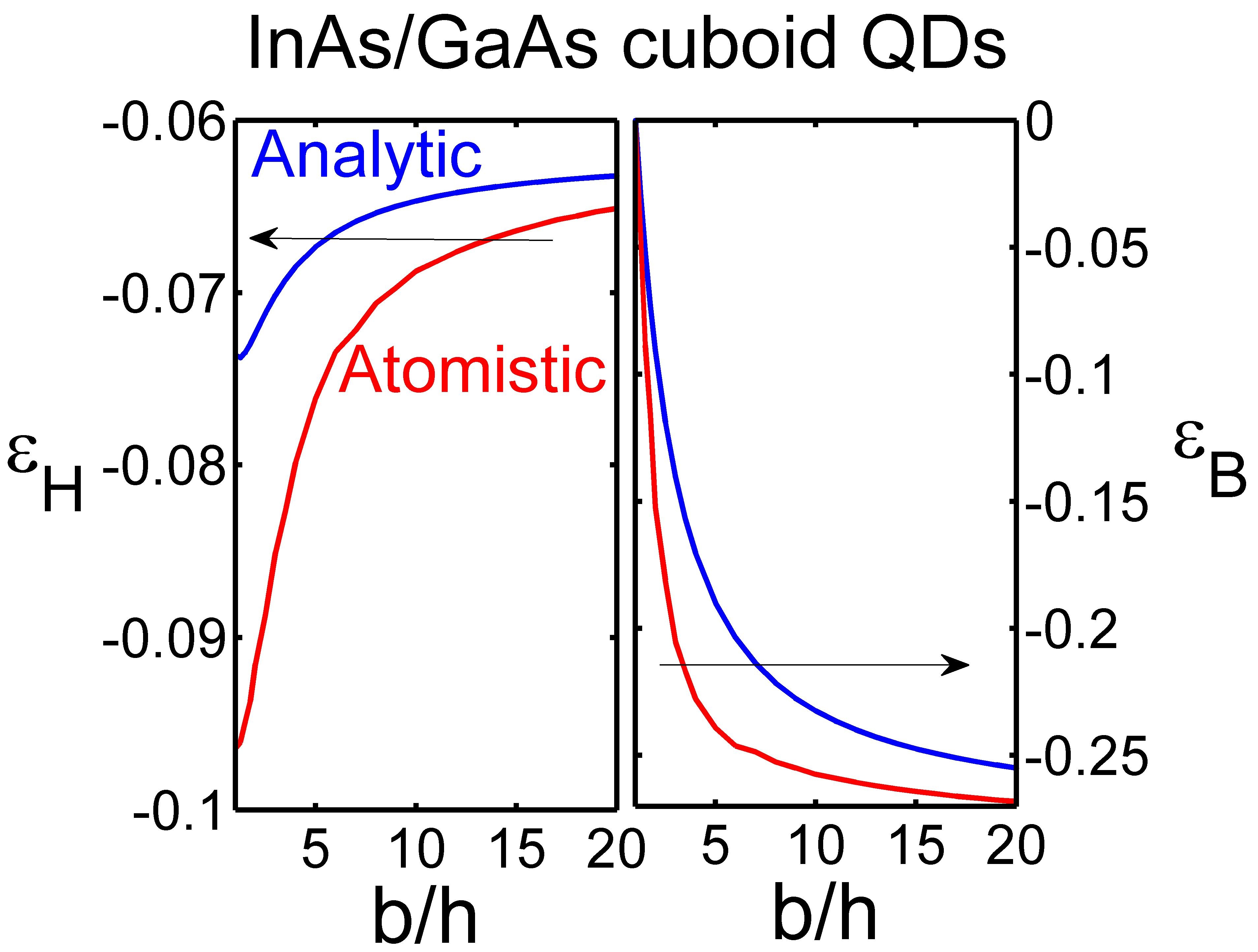}}
        \subfigure[]{\label{fig:b}\includegraphics[width=53mm]{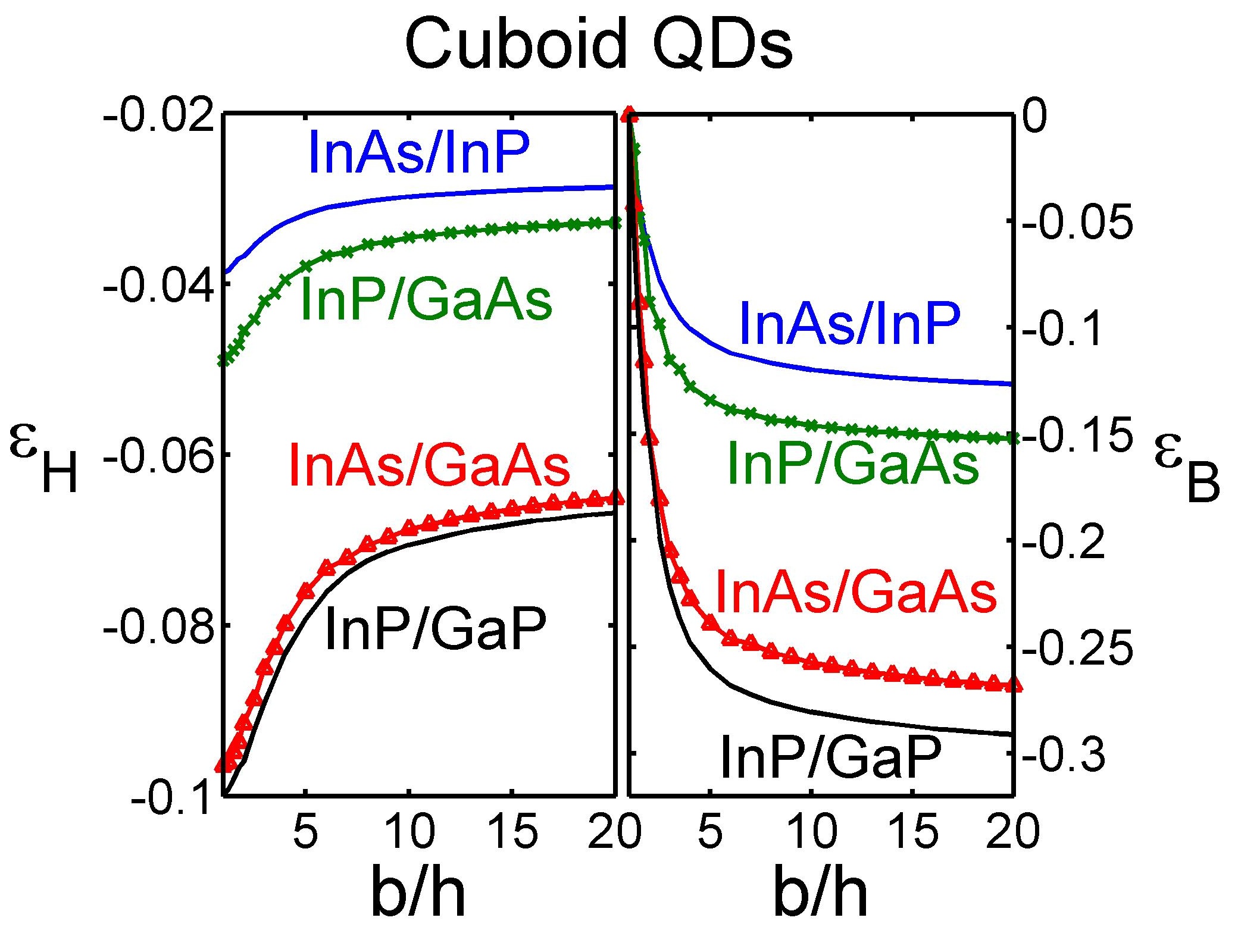}}
        \subfigure[]{\label{fig:c}\includegraphics[width=53mm]{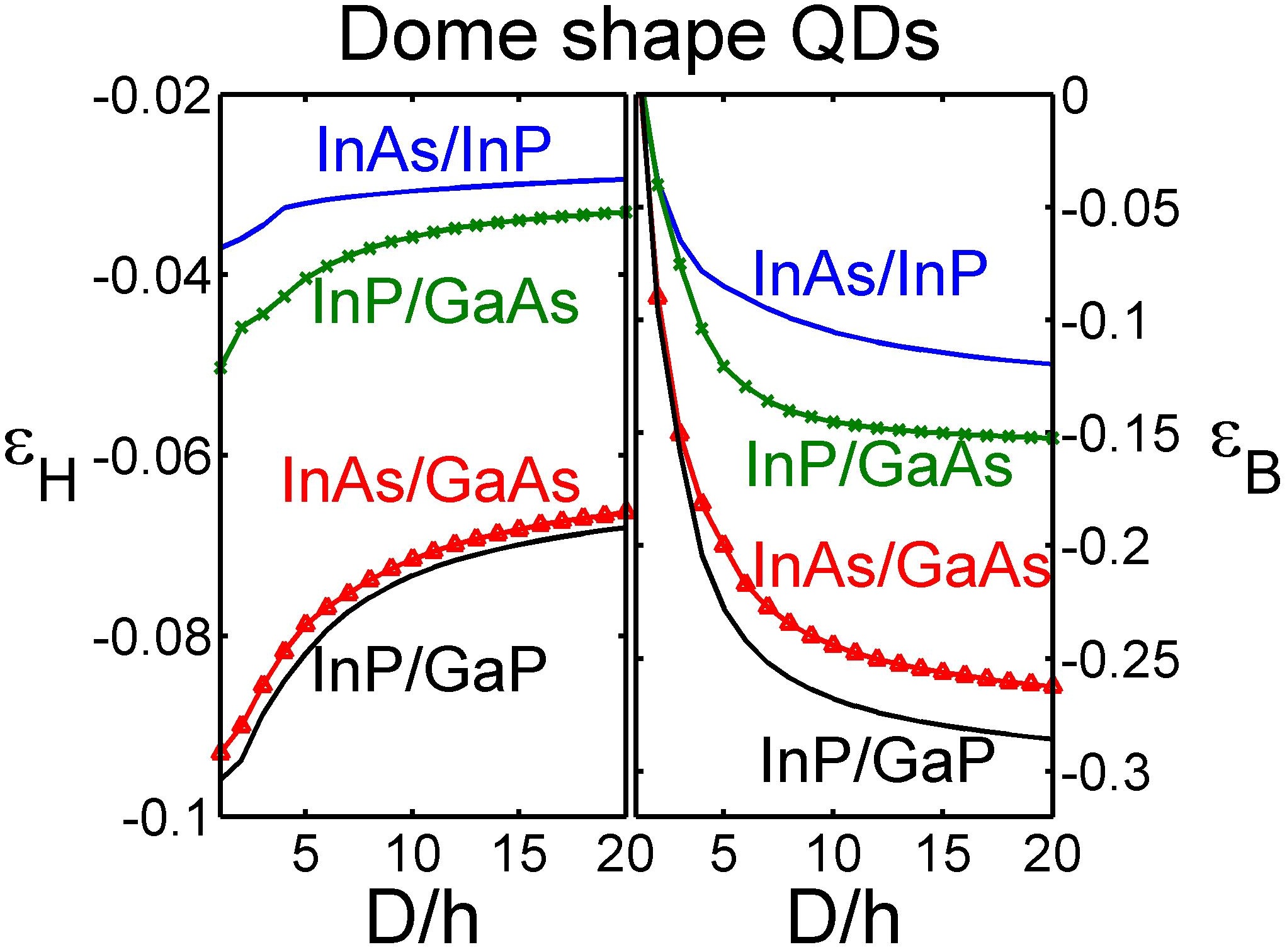}}                
\caption{a) The atomistic and analytic $\varepsilon_{H}$ and $\varepsilon_{B}$ at different base to hight ratio (b/h) of cuboid InAs/GaAs QD. The analytic solution underestimates the strain magnitudes. The atomistic $\varepsilon_{H}$ and $\varepsilon_{B}$ with different materials in cuboid (b) and dome shaped (c) QDs as a function of aspect ratios. The strain depends on the aspect ratio (b/h or D/h) and not on individual dimensions.}
\label{EHB_fig}
\end{figure}

Figure \ref{EHB_fig}a depicts the atomistic $\varepsilon_{H}$ and $\varepsilon_{B}$ versus the analytic solution in the middle of a cuboid quantum dot. The atomistic results show similar behavior to the analytic solution qualitatively but with significantly higher magnitudes. Similar trends exist in the other III-V material systems and shapes, as shown in Figs. \ref{EHB_fig}b and \ref{EHB_fig}c, respectively. The universal behavior that the atomistic strain in different materials depends on the aspect ratio of the QDs independent of actual physical dimensions can be captured with compact equations and be used to approximate strain in any QD, as discussed in the next section.
%

\section{Universal Behavior and its Applications}

As described in the previous section, atomistic strain depends only on the material properties and the aspect ratio of the QD. This behavior can be captured with compact equations as follows; First a function of aspect ratio is defined. This function is called \emph{universal behavior function} $U(AR)$, where the aspect ratio $AR$ is defined as $AR=\frac{b}{h}$ in cuboid QDs (the definition of $AR$ in dome shaped QDs is discussed later). The strain in the middle of the QD can be written in terms of this function:
\begin{equation}
\varepsilon(AR) =\varepsilon(AR \to \infty )+U(AR)~\left[ \varepsilon( AR=1)-\varepsilon(AR\to\infty) \right],
\label{eqn_UB1}
\end{equation}
\begin{equation}
U(AR) =\frac{\varepsilon(AR)-\varepsilon(AR\to\infty)}{\varepsilon( AR=1)-\varepsilon(AR\to\infty)},
\label{eqn_UB2}
\end{equation}

\begin{figure}[H]
\centering
        \subfigure[]{\label{fig:a}\includegraphics[width=60mm]{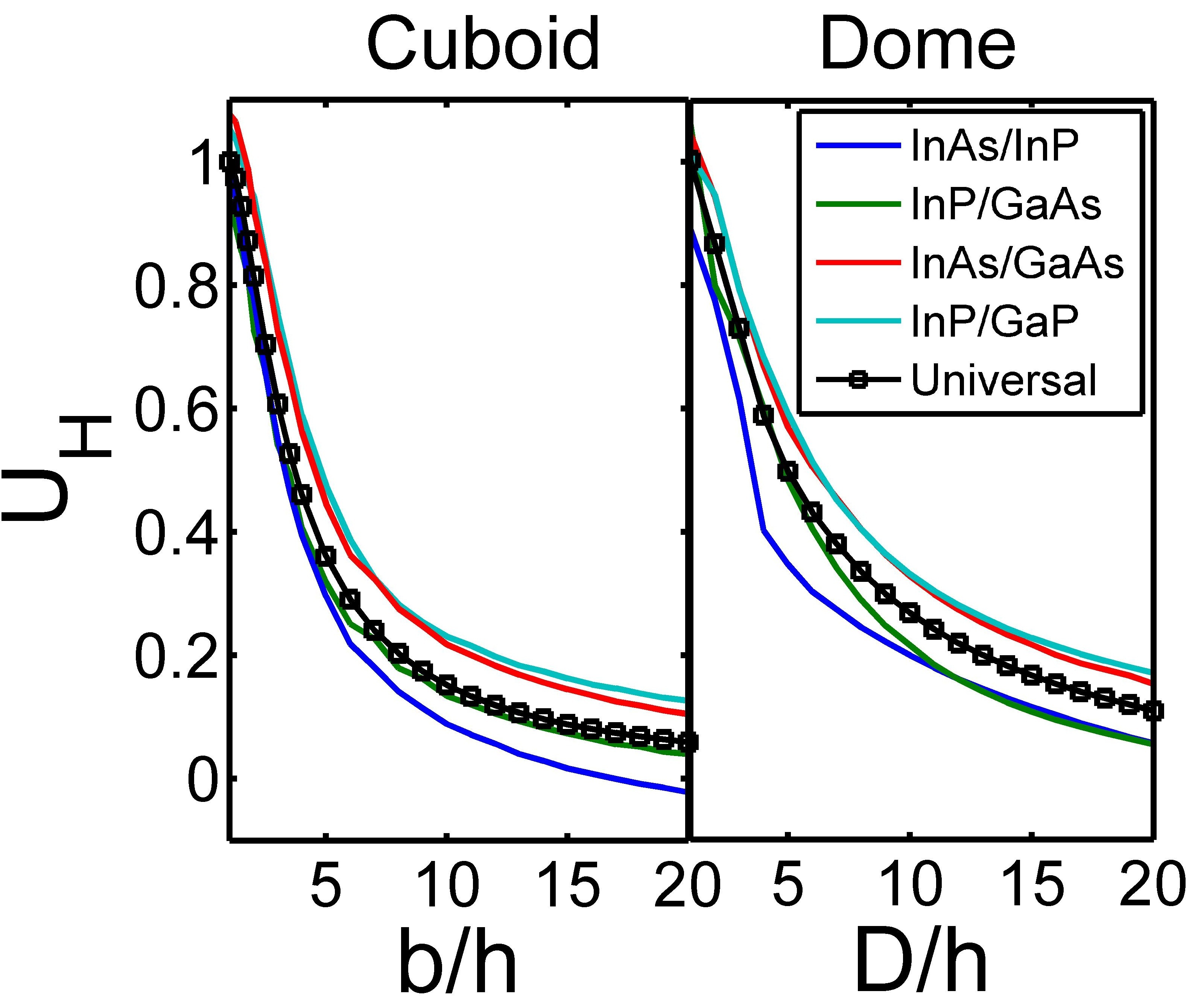}}
        \subfigure[]{\label{fig:b}\includegraphics[width=60mm]{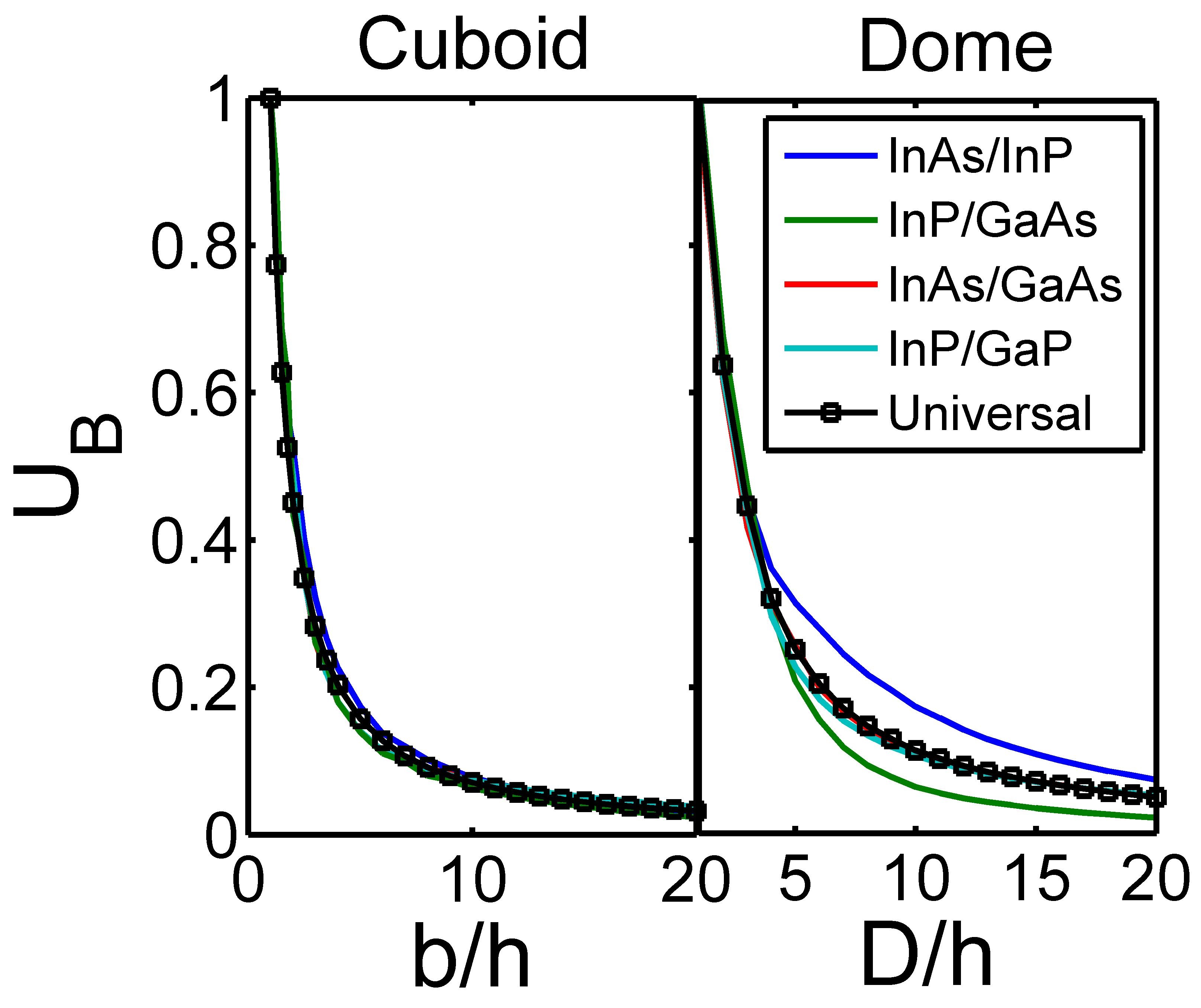}}
\caption{a) The hydrostatic $U_H(AR)$ and b) biaxial $U_B(AR)$ universal behavior data calculated for different materials in cuboid and dome shaped quantum dots. The black lines are obtained from equations (\ref{eqn_UB6}) and (\ref{eqn_UB7}). }
\label{UBF_fig}
\end{figure}

The universal function $U(AR)$ is different for hydrostatic (i.e. $U_H\left(AR\right)$) and biaxial strains (i.e. $U_B\left(AR\right)$), but both of them satisfy the following conditions:
\begin{equation}
U(AR) = \left\{
  \begin{array}{l l}
    0,~~AR\to\infty\\
    1,~~AR=1~
  \end{array} \right.
\label{eqn_UB3}
\end{equation}
Assuming that the strain values are known at the limits of $AR=1$ and $AR\to\infty$, the functions $U_H\left(AR\right)$ and  $U_B\left(AR\right)$ can be fitted numerically to the atomistic simulation results using equation (\ref{eqn_UB2}) as follows:
\begin{equation}
{{U}_{H}}\left( AR \right)=\frac{1}{0.75\text{ }\!\!~\!\!\text{ }{{\left( AR \right)}^{-0.4}}+0.25\text{ }\!\!~\!\!\text{ }{{\left( AR \right)}^{1.4}}},
\label{eqn_UB6}
\end{equation}
\begin{equation}
{{U}_{B}}\left( AR \right)=\frac{1}{{{\left( AR \right)}^{1.15}}}.
\label{eqn_UB7}
\end{equation}
To obtain a universal behavior function, the QDs with different materials and shapes need to fit into one equation which is the case as shown in Figs. \ref{UBF_fig}a and \ref{UBF_fig}b. 

The strain values at the limits of $AR=1$ and $AR\to\infty$ need to be determined still. The strain values are known for high aspect ratios ($AR\to\infty$) since in that regime the QD becomes a quantum well with accurate analytic strain solution: $\varepsilon_{H}\left(AR\to\infty \right)=-2 \varepsilon_0 \frac{1-2\nu_d}{1-\nu_d}$ and $\varepsilon_{B}\left(AR\to\infty \right)=-2 \varepsilon_0 \frac{1+\nu_d}{1-\nu_d}$, where $\varepsilon_0$ is the lattice misfit, $\varepsilon_0 = \frac{a_{dot}-a_{substrate}}{a_{dot}}$, and $\nu_d$ is the Poisson ratio of the dot's material.
On the other hand, in the case of $AR=1$, there is no analytic solution available for the strain except for $\varepsilon_{B}$ in cuboid QDs which is zero due to the symmetry: $\varepsilon_{B}\left(AR=1\right)=\varepsilon_{xx}+\varepsilon_{yy}-2\varepsilon_{zz}=0$, the rest of the values are approximated. In the cuboid QDs, the atomistic simulations reveal that $\varepsilon_{H}$ is close to $\varepsilon_{H}^{Cuboid}\left(AR=1\right)=-2.77 \varepsilon_0 \frac{1-2\nu_{av}}{1-\nu_{av}}$, where $\nu_{av}$ is the average Poisson ratio of the dot and the substrate materials. Whereas in the dome shaped QDs, the atomistic strain results are close to $\varepsilon_{H}^{Dome}\left(\frac{D}{h}=1\right)=-2.7 \varepsilon_0 \frac{1-2\nu_{av}}{1-\nu_{av}}$ and $\varepsilon_{B}^{Dome}\left(\frac{D}{h}=1\right)=0.2 \varepsilon_0 \frac{1+\nu_{av}}{1-\nu_{av}}$. Combining the strain values at the limits of $AR=1$ and $AR\to\infty$ with equation (\ref{eqn_UB1}) provides the strain equations in the cuboid and dome QDs:
\begin{equation}
\varepsilon_H^{Cuboid}(AR) = -2 \varepsilon_0 \frac{1-2\nu_d}{1-\nu_d}+U_H(AR)~\left[ -2.77 \varepsilon_0 \frac{1-2\nu_{av}}{1-\nu_{av}}+2 \varepsilon_0 \frac{1-2\nu_d}{1-\nu_d}) \right],
\label{eqn_UB4}
\end{equation}
\begin{equation}
\varepsilon_B^{Cuboid}(AR) = -2 \varepsilon_0 \frac{1+\nu_d}{1-\nu_d} +U_B(AR)\left[2 \varepsilon_0 \frac{1+\nu_d}{1-\nu_d} \right].
\label{eqn_UB5}
\end{equation}
\begin{equation}
\varepsilon_H^{Dome}(AR) = -2 \varepsilon_0 \frac{1-2\nu_d}{1-\nu_d}+U_H(AR) \left[ -2.7 \varepsilon_0 \frac{1-2\nu_{av}}{1-\nu_{av}}+2 \varepsilon_0 \frac{1-2\nu_d}{1-\nu_d}) \right],
\label{eqn_UB4_}
\end{equation}
\begin{equation}
\varepsilon_B^{Dome}(AR) = -2 \varepsilon_0 \frac{1+\nu_d}{1-\nu_d} +U_B(AR) \left[0.2 \varepsilon_0 \frac{1+\nu_{av}}{1-\nu_{av}} + 2 \varepsilon_0 \frac{1+\nu_d}{1-\nu_d} \right].
\label{eqn_UB5_}
\end{equation}

The exact atomistic strain values in QDs with different materials show  deviations less than 10$\%$ from the proposed universal function (shown in Fig. \ref{UBF_fig}). The black solid lines are obtained from the equations (\ref{eqn_UB4}) to (\ref{eqn_UB5_}). Defining $AR$ in the dome shaped QDs as $AR=0.85\frac{\sqrt{\pi}}{2}\frac{D}{h}$ makes it possible to use the same universal function in cuboid and dome shaped QDs.

Equations (\ref{eqn_UB6}) to (\ref{eqn_UB5_}) can be used to estimate the atomistic strain in cuboid and dome shaped quantum dots with good accuracy. These expressions can guide researchers studying quantum dots to understand the trends of the atomistic strain in self assembled QDs. In addition, the optical and electronic properties of QDs depends mainly on the strain. The effective mass and band edges of QDs can be determined from the strain values. Moreover, a prior knowledge of strain inside QDs helps to design and simulate QDs with required optical absorption spectrum. 
The  effect of strain on the band edges can be calculated using deformation potential theory\cite{bir1974symmetry}:
\[\Delta {{E}_{c}}={{a}_{c}}{{\varepsilon }_{H}}_{{}},\]
\[\Delta {{E}_{vHH}}={{a}_{v}}{{\varepsilon }_{H}}+\frac{b}{2}{{\varepsilon }_{B}}_{{}},\]
\[\Delta {{E}_{vLH}}={{a}_{v}}{{\varepsilon }_{H}}-\frac{b}{2}{{\varepsilon }_{B}}_{{}}, \]
where $\Delta {{E}_{c}}$ is the shift in the conduction band edge due to the strain, $\Delta {{E}_{vHH}}$ and  $\Delta {{E}_{vLH}}$ are the shifts in the heavy and light hole band edges, respectively. $a_c$, $a_v$, and $b$ are the deformation potential coefficients of the material.

\begin{figure}[H]
\centering
\includegraphics[width=70mm]{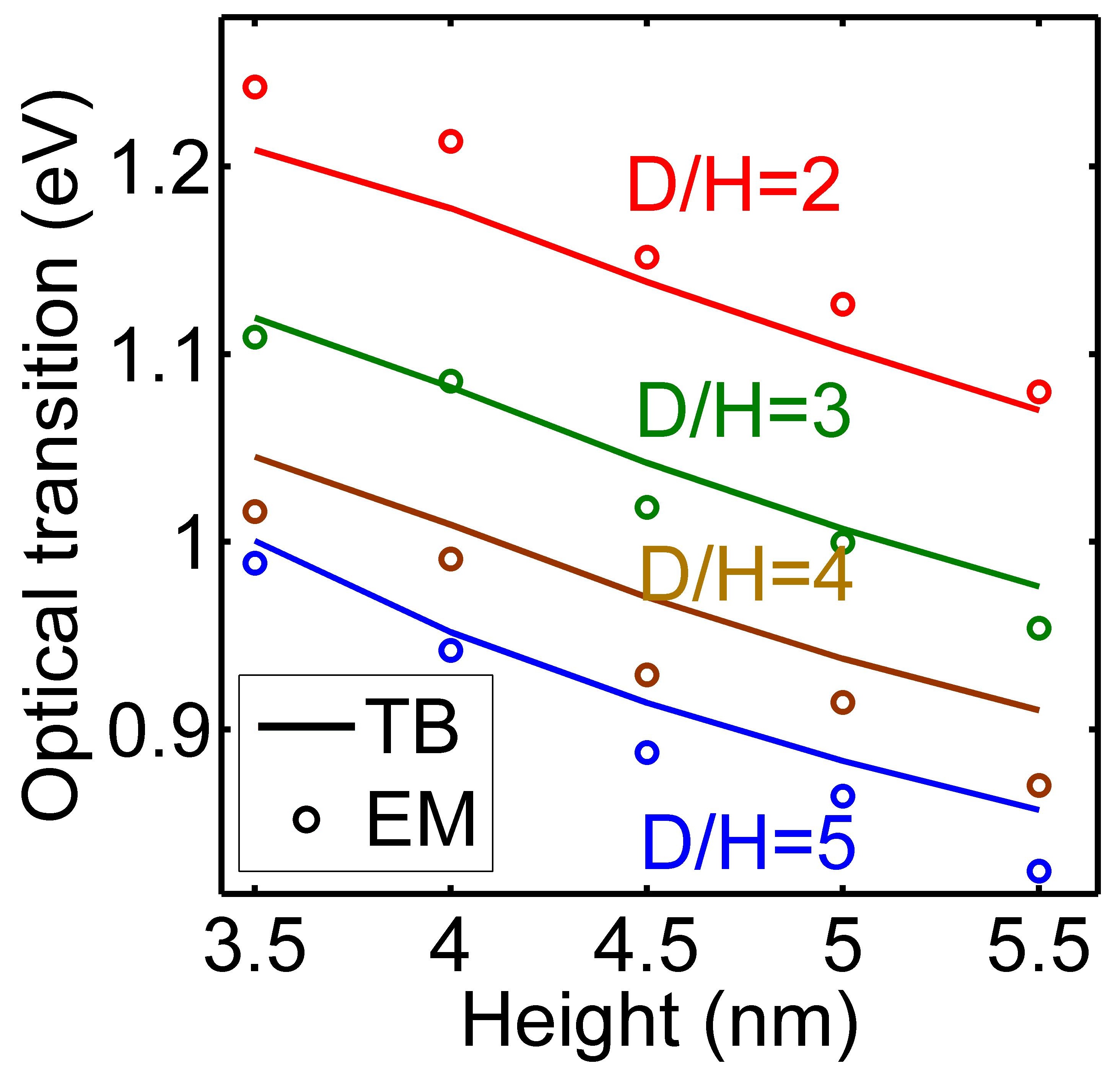}
\caption{Optical transitions in InAs/GaAs QDs as a function of the dot height for different AR calculated by the effective mass model (symbols) including the universal strain versus the full band tight binding calculations (solid lines). The deviation does not exceed 4.5\%.}
\label{Optical_transition_fig}
\end{figure}

To show the impact of these expressions, the optical transition energies of an InAs/GaAs dome shaped quantum dot with height of 5 nm and diameter of 20 nm are calculated here using different techniques and compared against experimental measurements\cite{tate}. The effective mass simulation was performed based on an efficient scheme \cite{Tarek}. Including the universal strain behavior in the effective mass model (EM) reduces the error in the predicted optical transition from 25\% to 4\% as shown in Table \ref{Table}. Despite the error not being as small as a full band tight binding (TB) simulation (1.6\%), the computational cost is much less as indicated in Table \ref{Table}. The tight binding simulations were performed in NEMO5 using $sp^3d^5s^*$ basis \cite{nemo5}. The values of the InAs effective masses under strain are taken from \cite{klimeck2002development}. Figure \ref{Optical_transition_fig} shows the optical transitions calculated from the effective mass model including the universal strain and the full band tight binding simulation. The deviation of the results between these methods does not exceed 4.5\%.

\begin{table}[H]
\centering
\begin{tabular} { |c|c|c|c|c| }
  \hline
  ~ & EM no Strain & EM with Strain & TB Atomistic & Experimental \\ \hline
   Optical transition & 0.7169 eV & 0.9145 & 0.9377 & 0.976 - 0.93 eV  \\ \hline
   Error (vs 0.953 eV) & 25\% & 4\% & 1.6\% & ~  \\
   Number of CPUs & 1 & 1 & 216 & ~ \\
   Time/CPU (s) & 80 & 81 & 7856 & ~ \\
  \hline
\end{tabular}
\caption{Calculated optical transition energy using different techniques versus the experimental measurement. Using the universal behavior equations to include the effect of strain on the band edges and effective masses significantly improves the effective mass model results without any additional computational cost.}
\label{Table}
\end{table}

%

\section{Conclusion}

Atomistic simulations are known to be the most accurate tool for studying QDs, however they are computationally expensive. On the other hand, the analytic continuum solutions are very fast, but they are based on the assumptions that significantly reduce the accuracy of the predicted strain in QDs. Using the fact that atomistic strain in QDs depends on the aspect ratio of the QD (\emph{universal behavior}) a novel method to predict the strain is proposed. It is shown that this universal behavior of atomistic strain exists in QDs with different shapes (e.g. cuboid and dome shaped QDs) and materials (e.g. InAs/GaAs, InP/GaP, etc.). The proposed compact strain equations can be used to design the optical properties of the QDs or predict the dimensions of a fabricated self assembled QD from its optical transitions with good accuracy. 

\section{Acknowledgment}
This research is also part of the Blue Waters sustained-petascale computing project, which is supported by the National Science Foundation (award number ACI 1238993) and the state of Illinois. Blue Waters is a joint effort of the University of Illinois at Urbana-Champaign and its National Center for Supercomputing Applications. This work is also part of the "Accelerating Nano-scale Transistor Innovation with NEMO5 on Blue Waters" PRAC allocation support by the National Science Foundation (award number OCI-0832623). 

\bibliographystyle{ieeetr}
\bibliography{thesis}

\section*{Appendix: Analytical solution for strain in quantum dots}
In this section, we have derived simple analytical expressions that describe strain in self assembled quantum dots based on continuum elasticity theory. 
The derivation here uses the analytical solution for strain inside the quantum dot provided in \cite{pearson2000analytical}. The reader is advised to read \cite{pearson2000analytical} first. 
Before we derive the general expressions, it is helpful to understand the behavior of the strain in  an ideal quantum well first.

\subsection*{Cuboid quantum dot}
In \cite{pearson2000analytical}, an analytical expression for the strain of a clipped pyramidal quantum dot has been obtained using the continuum elasticity and Green's function technique. For notations and the meaning of the parameters used in the following derivation, please refer to the mentioned paper and its predecessors\cite{pearson2000analytical,downes1997simple,faux1996simple,downes1995calculation}.
We will follow the same assumptions of the paper \cite{pearson2000analytical} except for assuming constant hydrostatic strain. Henceforth the assumptions of the proposed analytical solution are:
\begin{itemize}
  \item The solution is based on a stress field Green's function that has been introduced for quantum wires\cite{downes1995calculation}, for which at least one strain component should be known. This is suitable for quantum wires and quantum wells but not accurate for quantum dots. This assumption is the main reason behind deviation from the atomistic result of the quantum dot as shown in section II. The matching of the analytical solution with the atomistic results at high values of base to hight ratio happens because the quantum dot becomes more like a quantum well. 
  \item Continuous, elastic, linear, and isotropic mediums.
  \item Infinite substrate.
  \item Elasticity constants of dot and substrate are the same.
  \item The quantum dot is a cuboid with base dimensions $a=b$ and height $H$.
  \item The strain is calculated at a point in the middle of the quantum dot.
\end{itemize}
The hydrostatic strain $ \varepsilon_{H}$ is defined as,

\[ \varepsilon_{H}  = \varepsilon_{xx}+\varepsilon_{yy}+\varepsilon_{zz} ,\]
\[ \varepsilon_{H}  = \frac{1}{E} \left(  \sigma_{xx}-\nu\left( \sigma_{yy}+\sigma_{zz}\right) + \sigma_{yy}-\nu\left( \sigma_{zz}+\sigma_{xx}\right) +\sigma_{zz}-\nu\left( \sigma_{xx}+\sigma_{yy}\right) \right),\]
\[ \varepsilon_{H}  = \frac{1 - 2 \nu}{E} \left( \sigma_{xx}+\sigma_{yy}+\sigma_{zz}   \right), \]
\begin{equation}
 \varepsilon_{H}  = \frac{1 - 2 \nu}{E} \sigma_{h}.
\label{eqn_App_1}
\end{equation}

The biaxial strain $\varepsilon_{B}$ is defined as,
\[ \varepsilon_{B}  = \varepsilon_{xx}+\varepsilon_{yy}- 2 \varepsilon_{zz} ,\]
in a similar way to the hydrostatic strain,
\[ \varepsilon_{B}  = \frac{1 + \nu}{E} \left( \sigma_{xx}+\sigma_{yy}-2\sigma_{zz}   \right), \]
\begin{equation}
 \varepsilon_{B}  = \frac{1 + \nu}{E} \sigma_{b}.
\label{eqn_App_2}
\end{equation}

Following the expressions in \cite{pearson2000analytical} and using the assumptions mentioned before:
The dot has a cuboid shape,
\[h \rightarrow \infty ~~,~~ f \rightarrow 0 ~~,~~ H = hf,\]
since we are calculating the strain in the middle of the dot then,
\[x_1=x_2=0~~, x_3=\frac{H}{2},\]
calculating the parameters in \cite{pearson2000analytical} under these assumptions with some algebraic simplifications yields,
\[ X_1 = X_2 = X_3 = p a ~~, p= \pm 1, \]
\[ Y_1 = Y_2 = Y_3 = q b ~~, q= \pm 1, \]
\[ Z_2 = \frac{-H}{2} ~~ , Z_3 = \frac{H}{2}, \]
\[ A = B = H_1 =0 ~~ , h H_1 = 1, \]
\[ S_2=S_3=S=\sqrt{ a^2 + b^2 + \frac{H^2}{4}}, \]
\[ C_{11}= a^2 + b^2 + q b S \]
\[ D_{11} = pa Z_n\]

\begin{equation}
\alpha  = \frac{p a (S^2 +q b S)}{a^2 + \frac{H^2}{4}},
\label{eqn_App_3}
\end{equation}
\begin{equation}
\beta  = \frac{p Z_n (b^2 +q b S)}{a^2 + \frac{H^2}{4}},
\label{eqn_App_4}
\end{equation}
\begin{equation}
\frac{\beta}{\alpha} = (-1)^{n+1}\frac{H}{2a} \frac{(b^2 +q b S)}{(S^2 +q b S)}.
\label{eqn_App_5}
\end{equation}
 
Since $H1=0$ and $h H1=1$ , the expression for $\sigma_{11}$ is simplified to only one term in the summation 
\begin{equation}
\sigma_{11}  = - 4 \pi \Lambda + \Lambda \sum_{\substack{
   p = \pm 1 \\
   q = \pm 1 \\
   n = 2,3
  }}
 {(-1)^{n+1} q ~ tan^{-1}(\frac{\beta}{\alpha})},
\label{eqn_App_6}
\end{equation}
where $\Lambda = \frac{\varepsilon_0 E}{4\pi (1-\nu)}$, substituting from equation (\ref{eqn_App_5}) and simplifying the summation,

\begin{equation}
\sigma_{11}  = - 4 \pi \Lambda + 4 \Lambda \left( tan^{-1}\left(\frac {H (b^2+ b S)}{2a (S^2+b S)} \right)-tan^{-1}\left(\frac {H (b^2- b S)}{2a (S^2 - b S)} \right) \right).
\label{eqn_App_7}
\end{equation}

For $\sigma_{xx}$ and $\sigma_{yy}$, put $a=b$ which is the base length, and put $H=h$ which is the height of the dot in equation (\ref{eqn_App_7}),

\begin{multline}
\sigma_{xx} = \sigma_{yy}  = - 4 \pi \Lambda + 4 \Lambda tan^{-1}\left(\frac{1}{2}\frac {\frac{b}{h}+\sqrt{2\left(\frac{b}{h}\right)^2 +\frac{1}{4}}}{2\left(\frac{b}{h}\right)^2 +\frac{1}{4} + \frac{b}{h} \sqrt{2\left(\frac{b}{h}\right)^2 +\frac{1}{4}}} \right) \\ - 4 \Lambda tan^{-1}\left(\frac{1}{2}\frac {\frac{b}{h}-\sqrt{2\left(\frac{b}{h}\right)^2 +\frac{1}{4}}}{2\left(\frac{b}{h}\right)^2 +\frac{1}{4} - \frac{b}{h} \sqrt{2\left(\frac{b}{h}\right)^2 +\frac{1}{4}}} \right), 
\label{eqn_App_8}
\end{multline}

with some algebraic and trigonometric  simplifications, one obtains

\begin{equation}
\sigma_{xx} = \sigma_{yy}  = - 4 \pi \Lambda + 8 \Lambda tan^{-1}\left( \frac{1}{2\sqrt{\left(\frac{b}{h}\right)^2+\frac{1}{4}}} \right).
\label{eqn_App_9}
\end{equation}

For $\sigma_{zz}$, exchange $a$ and $H$ in equation (\ref{eqn_App_7}) then put $a=b$ and $H=h$,

\begin{multline}
\sigma_{zz}  = - 4 \pi \Lambda + 4 \Lambda tan^{-1}\left( \frac{\left(\frac{b}{h}\right)^2}{2} \frac{\frac{b}{h}+\sqrt{1+\frac{5}{4}\left(\frac{b}{h}\right)^2}}{1+\frac{5}{4}\left(\frac{b}{h}\right)^2+\frac{b}{h}\sqrt{1+\frac{5}{4}\left(\frac{b}{h}\right)^2}} \right) \\ - 4 \Lambda tan^{-1}\left( \frac{\left(\frac{b}{h}\right)^2}{2} \frac{\frac{b}{h}-\sqrt{1+\frac{5}{4}\left(\frac{b}{h}\right)^2}}{1+\frac{5}{4}\left(\frac{b}{h}\right)^2-\frac{b}{h}\sqrt{1+\frac{5}{4}\left(\frac{b}{h}\right)^2}} \right) . 
\label{eqn_App_10}
\end{multline}

with some algebraic and trigonometric  simplifications, one obtains

\begin{equation}
\sigma_{zz}  = - 4 \pi \Lambda + 8 \Lambda cot^{-1}\left( \frac{2\sqrt{1+\frac{5}{4}\left(\frac{b}{h}\right)^2}}{\left(\frac{b}{h}\right)^2} \right).
\label{eqn_App_11}
\end{equation}

Substituting equations (\ref{eqn_App_9}) and (\ref{eqn_App_11}) into equations (\ref{eqn_App_1}) and (\ref{eqn_App_2}), we get the final expressions for the hydrostatic and biaxial strain:

\begin{equation}
 \varepsilon_{H}  = \frac{1 - 2 \nu}{E} \left( - 12 \pi \Lambda + 16 \Lambda tan^{-1}\left( \frac{1}{2\sqrt{\left(\frac{b}{h}\right)^2+\frac{1}{4}}} \right) + 8 \Lambda cot^{-1}\left( \frac{2\sqrt{1+\frac{5}{4}\left(\frac{b}{h}\right)^2}}{\left(\frac{b}{h}\right)^2} \right)\right),
\label{eqn_App_12}
\end{equation}

\begin{equation}
 \varepsilon_{B}  = \frac{1 + \nu}{E} 16 \Lambda \left( tan^{-1}\left( \frac{1}{2\sqrt{\left(\frac{b}{h}\right)^2+\frac{1}{4}}} \right) - cot^{-1}\left( \frac{2\sqrt{1+\frac{5}{4}\left(\frac{b}{h}\right)^2}}{\left(\frac{b}{h}\right)^2} \right) \right).
\label{eqn_App_13}
\end{equation}

\end{document}